# Peak intrinsic thermal conductivity in non-metallic solids and new interpretation of experimental data for argon

Ahmed Hamed[1], Anter El-Azab[2]

[1]School of Nuclear Engineering, Purdue University, West Lafayette, Indiana 47907, USA

[2]School of Materials Engineering, Purdue University, West Lafayette, Indiana 47907, USA

Abstract

The inelastic nature of 3-phonon processes is investigated within the framework of perturbation theory and linearized Boltzmann Transport Equation. By considering the energy conservation rule governing this type of interactions in a statistical average sense, the impact of different forms of the regularized energy-conserving Dirac delta function on 3-phonon scattering rates was evaluated. Strikingly, adopting Lorentz distribution, in accordance with the shape of eigenenergy broadening of phonon normal modes due to the leading term of crystal anharmonicity, was found to play a critical role in activating umklapp processes at low temperature, leading to intrinsic lattice thermal conductivity peak at finite temperature for perfect crystal. This characteristic behavior, unique to the Lorentzian, lays foundation for developing adjustable-parameter-free computational models for reliable prediction of the finite lattice conductivity at low temperature, even in the absence of extrinsic scattering processes (e.g., by crystal imperfections and boundary). An iterative solution scheme for Boltzmann Transport Equation was used to compute the intrinsic thermal conductivity of solid argon over the entire temperature range (2 – 80 K). For the first time, the experimentally observed $T^2$ behavior in the low temperature, T, limit and the peak temperature (~ 8 K) were successfully recovered, in addition to the classical high temperature $T^{-1}$ behavior above 20 K by the sole use of 3-phonon processes. The good agreement with experiment indicates that phonon-phonon interactions dominate over the entire temperature range in argon, contrary to previous hypotheses that the sub-peak regime is dominated by phonon-defect scattering. Anisotropy in thermal conductivity of single crystal at low temperature due to phonon focusing was observed. In addition, argon conductivity is underestimated by an order of magnitude in Single Mode Relaxation Time approximation, where the collective nature of phonon mode relaxation is ignored.

## I. INTRODUCTION

The current study attempts to quantitatively assess the role played by 3-phonon scattering processes in thermal resistivity of pristine crystals at low temperature. Accurate representation of phonon intrinsic scattering rates at low temperature will pave the way for developing high fidelity models for thermal conductivity predictions at all temperatures. This is both of fundamental interest and technical importance in applications aiming to tailor the thermal performance of materials by tuning phonon processes via introducing nanostructured features to address thermal management challenges, e.g., enhancing thermal dissipation in nanoelectronic devices and improving thermal insulation to increase the efficiency of thermoelectrics.[1,2]

Debye exclusively ascribed the finite lattice thermal conductivity of perfect crystal to the anharmonic part of crystal energy, since harmonic crystal would have infinite conductivity.[3] Peierls elaborated on that and pointed out the role played by the discrete nature of the lattice, and attributed the intrinsic lattice resistivity to what he called umklapp processes.[3-9] Moreover, Peierls predicted qualitatively an exponential increase in the thermal conductivity by approaching 0 K, as these processes get frozen out.[4-6] In dielectric crystals, the contradiction between this expectation and the experimental observations at very low temperature is resolved by considering phonon scattering by extrinsic resistive processes (e.g., lattice imperfections and crystal boundary), since phonons are the main heat carriers in these materials. Although this view proved to be successful in reproducing experimental results and estimating the levels of defect concentrations, it leaves the fundamental question about the finite, large conductivity of perfect crystal at 0 K unanswered. In some common phonon based conductivity models normal processes are treated as non-resistive processes (valid only for Debye approximation), while in others the role of normal processes is completely ignored.[7-9] Given the fact that the classification of individual 3-phonon scattering events as normal or umklapp processes depends on how the primitive cell in the reciprocal space is chosen,[8] the interplay between these two mechanisms should be considered for accurate prediction of 3-phonon scattering rates.

Another challenge usually encountered in the evaluation of 3-phonon scattering rates within the framework of harmonic approximation—perturbation theory is the handling of Dirac delta function appearing in the Fermi golden rule[7]. Although 3-phonon scattering is inelastic, it has been always treated as elastic interaction[7]. In this communication, we seek a more rigorous consideration of the inelastic nature of 3-phonon scattering, within the standard perturbation theory approach, by applying energy conservation rule (which governs 3-phonon scattering processes) in a statistical average sense. In this regard, we show the profound impact of the statistical consideration of energy conservation rule on low temperature intrinsic lattice thermal conductivity prediction using Boltzmann Transport Equation (BTE). The observed impact is mainly important at low temperatures, while its effect is masked at high temperatures. It turned out that the use of Lorentz distribution to represent the regularized Dirac delta function has a determining role in showing a peak at finite temperature in the intrinsic lattice thermal conductivity and its decay back toward zero as the temperature approaches 0 K. The credibility of the Lorentzian representation, as compared to other statistical distribution such as the Gaussian, to capture the inelastic nature of 3-phonon processes is substantiated by experiment and theory. Recall that, in conformity with the theory of forced resonance, perturbation theory prediction of eigenenergy broadening due to three-phonon processes alone follows Lorentz distribution, which is in agreement with the experimentally observed phonon lineshape[10,11]. Moreover, the deviation of the lineshape from the Lorentzian indicates the contribution of other phonon scattering mechanisms (including higher order phonon-phonon interaction processes) to the eigenenergy broadening[10]. Microscopically, this characteristic behavior of intrinsic thermal conductivity profile is a consequence of the fact that the use of Lorentz distribution brings out phonon-phonon interactions at low temperature as an additional resistive mechanism to thermal transport that should also be considered in conductivity prediction below peak conductivity temperature in dielectrics. To the best of our knowledge, this is the first computational study that predicts this low temperature behavior of thermal conductivity by the sole use of 3-phonon processes and Fermi golden rule for solving the linearized form of BTE. We believe that such a finding was not captured in previous models due to a lack of systematic scrutiny thereof of the effect of the shape of Dirac delta distribution on the results reported in literature over the whole range of temperature. Although previous studies reported no difference in the calculated thermal conductivity when Lorentzian distribution was tested, as compared to other mathematical representations, the temperature range considered was way high above the peak conductivity temperature, e.g., Ref. [12]. By using perturbation-theory-based expression for phonon collision kernel and consulting phonon kinetic theory, we observed a conductivity peak at 8 K for argon. The overall conductivity profile strikingly indicates that phonon-phonon interactions are the dominant scattering mechanism over the entire temperature range for solid argon. Of course this finding of the dominance of phonon-phonon interactions at low temperature should not be in any way generalized to all dielectrics.

Argon was chosen for the current investigation due to its strong crystal anharmonicity even at 0 K, simple structure, high isotopic purity, and the existence of reasonably adequate classical potential to describe the atomic interactions. Thermal conductivity and vibrational properties of solid argon were subjects of several experimental[13-20] and numerical studies.[11,21-35] The experimentally observed $T^2$ behavior of thermal conductivity at low temperature (below the peak) indicates that grain boundary scattering (with theoretical prediction of $T^3$ dependence) is not the dominant scattering mechanism in this regime. In addition, orders of magnitude difference between the estimated phonon mean free path in this temperature range (about $10^{-3}$ mm at 2 K) and the average grain size of the examined specimens (~ mm) supports the above assertion.[18,19] Furthermore, the reported high purity of the samples used in the relevant experiments renders scattering by point defects as an explanation for this temperature dependent behavior of thermal conductivity implausible. The remaining active resistive mechanism was speculated to be scattering by dislocations. Gupta and Trikha[22,23] utilized a semi-empirical model to reproduce the experimental measurements, where they used dislocation density as an adjustable parameter. Crystal state of used specimens was not examined in any of the available experiments. In addition, White and Woods[17] reported on the absence of any significant changes in their thermal conductivity measurements when an annealed specimen was used. The latter reports cast doubt on the validity of the proposition of dislocations as the dominant scattering mechanism in the context of interpreting experimental data at temperatures below the thermal conductivity peak.

Christen and Pollack[9] inserted Holland's phenomenological expression for phonon relaxation times into Krumhansl model to calculate the thermal conductivity of argon at low temperature, and they ended up with a conclusion that an additional scattering term is required to recover experimental data. By applying lattice dynamics approach, Julian[26] was the first to invoke a mechanistic model for argon lattice thermal conductivity evaluation, using semi-empirical interatomic potentials to furnish the interatomic force constants required for calculation of the elements of 3-phonon processes transition rate matrix. The formulation was derived from perturbation theory, and BTE was solved by variational method. His study confirmed Peierl's prediction at low temperature. In addition, he strikingly found $T^{-1}$ temperature dependency to begin right above a one quarter of Debye temperature ~ 20 K (in contrast to the theoretical prediction for the onset of this behavior above Debye temperature). Many succeeding numerical studies exhibited same behavior in high temperature regime. Nevertheless, Krupskii and Manzhelii[16] and, very recently, Feng and Ruan[35] referred to the significant role played by four-phonon processes in the high temperature limit and used this mechanism to interpret their experimentally observed $T^{-2}$ temperature dependency. On the other hand, Clayton and Batchelder[19] attributed this deviation to thermal expansion and showed that under constant volume condition $T^{-1}$ pattern is retained.

At low temperature, the quantization of the vibrational energy and the wave nature of thermal transport should be considered. In situations where non-local effects are not important, the quasi-particle picture can be invoked, where wave effects are incorporated into the scattering term. Hence, thermal transport can be studied by tracking the temporal evolution of phonon population in phase space using the semi-classical phonon BTE. The scattering term, assuming small crystal anharmonicity, can be determined from perturbation theory,[35-37] which should be valid for temperatures up to roughly one third of the melting temperature.[10,36] However, the main criteria are the relative amplitude of atomic displacement with respect to interatomic spacing and the relative frequency shift and width with respect to harmonic frequency.[10,33]

In the current investigation, thermal conductivity has been predicted using the framework of perturbation theory and the linearized BTE, in which a real structure of phonon states for solid argon is incorporated. Utilizing the Lorentz distribution to represent the regularized Dirac delta function, our approach perfectly captures the temperature dependence behavior of thermal conductivity in both low ($T^2$) and high ($T^{-1}$) temperature limits, without the assumption of defect or grain boundary scattering. In addition, by the virtue of the iterative scheme, good agreement with experiment between 2 K and 80 K was met.

## II. THEORY AND COMPUTATIONAL APPROACH

By applying the principle of microscopic reversibility and linearization technique about thermodynamic equilibrium, we can express the scattering term of BTE in terms of the equilibrium distributions, $\bar{n}_{qs}$ (Bose-Einstein distribution with zero chemical potential) and transition rates $\bar{P}$ at temperature $T$, in addition to the function $\psi_{qs}$ measuring the deviation of phonon occupation number of a given mode at point $r$ and time $t$, $n_{qs}(r,t)$, from the equilibrium distribution, such that $n_{qs} \simeq \bar{n}_{qs} + \frac{1}{k_B T} \psi_{qs} \bar{n}_{qs} (\bar{n}_{qs} + 1)$. In steady state, by equating the drift term due to an applied temperature gradient, $\nabla T$, with the collision term, linearized BTE takes the form

$$\frac{\hbar \omega_{qs}}{k_B T^2} \upsilon_g^{qs} \cdot \nabla T \bar{n}_{qs} (\bar{n}_{qs} + 1) = -\frac{1}{k_B T} \sum_{q's',q''s''} [\bar{P}_{qs,q's'}^{q''s''} (\psi_{qs} + \psi_{q's'} - \psi_{q''s''}) + \frac{1}{2} \bar{P}_{qs}^{q's',q''s''} (\psi_{qs} - \psi_{q's'} - \psi_{q''s''})]. \quad (1)$$

In this expression, $\upsilon_g^{qs}$ denotes the group velocity of the phonon normal mode $qs$ (where, $q$ stands for the phonon wavevector and $s$ is the polarization branch index) with angular frequency $\omega_{qs}$, while $k_B$ and $\hbar$ stand for Boltzmann and the reduced Plank constants, respectively. The equilibrium transition rates for individual fusion $\bar{P}_{qs,q's'}^{q''s''}$ and fission $\bar{P}_{qs}^{q's',q''s''}$ events can be obtained using Fermi golden rule and the properties of phonon ladder operators. In long-wavelength limit, where continuum approximation is applicable, thermodynamic Grüneisen parameter $\gamma$ can be used alongside with the sound velocity $\bar{\upsilon}_s$ to give a measure of the crystal anharmonicity. Under long-wavelength approximation, only one mode-averaged parameter is used to represent crystal anharmonicity. Following Srivastava's notation,[7] the three-phonon scattering rates are given, respectively, by

$$\bar{P}_{qs,q's'}^{q"s"} = \frac{\pi\hbar\gamma^2}{\rho N_o \Omega \bar{\upsilon}_s^2} \omega_{qs}\omega_{q's'}\omega_{q"s"}\bar{n}_{qs}\bar{n}_{q's'}(\bar{n}_{q"s"}+1)\delta(\omega_{qs}+\omega_{q's'}-\omega_{q"s"})\delta_{q+q'+q",G}, \tag{2a}$$

and

$$\bar{P}_{qs}^{q's',q"s"} = \frac{\pi\hbar\gamma^2}{\rho N_o \Omega \bar{\upsilon}_s^2} \omega_{qs}\omega_{q's'}\omega_{q"s"}(\bar{n}_{qs}+1)\bar{n}_{q's'}\bar{n}_{q"s"}\delta(\omega_{qs}-\omega_{q's'}-\omega_{q"s"})\delta_{q-q'+q",G}. \tag{2b}$$

In the above, $N_o$ stands for the total number of the primitive unit cells, $\Omega$ for the primitive unit cell volume, $\rho$ for material density, $G$ for reciprocal lattice vector, $\delta_{q+q'+q",G}$ for Kronecker delta function, and $\delta(\omega_{qs}\pm\omega_{q's'}-\omega_{q"s"})$ is Dirac delta function; see Ref. [7] for more details. In harmonic lattice dynamics approach, phonon eigenfrequencies and eigenvectors for a given wavevector are found by diagonalizing the dynamical matrix that is constructed from the Fourier transformed harmonic force constants. This can be found directly from a given analytic interatomic potential or by applying frozen phonon method to MD or electronic structure simulations.[7,11,33-35] Under energy continuum approximation, on the other hand, dispersion curves in certain discrete directions of the reciprocal space replaces the eigenfrequencies of individual modes. They are generated from few points in each direction using curve fitting techniques. The frequencies of these points can be obtained from experiment or by solving the secular determinant.[11,33-36] Using Mie-Lenard-Jones (6-12)—all neighbors interatomic potential for argon with two fitting parameters that reproduce the lattice constants and sublimation energy at 0 K,[28] dispersion curves in the three high symmetry crystallographic directions [001], [110], and [111] are generated by applying trigonometric function fitting. These dispersion curves are shown in Fig. 1 for the three polarization branches. The three eigenvalues (for the three different polarization types) of selected points in each direction were found by solving the secular determinant of the dynamical matrix.

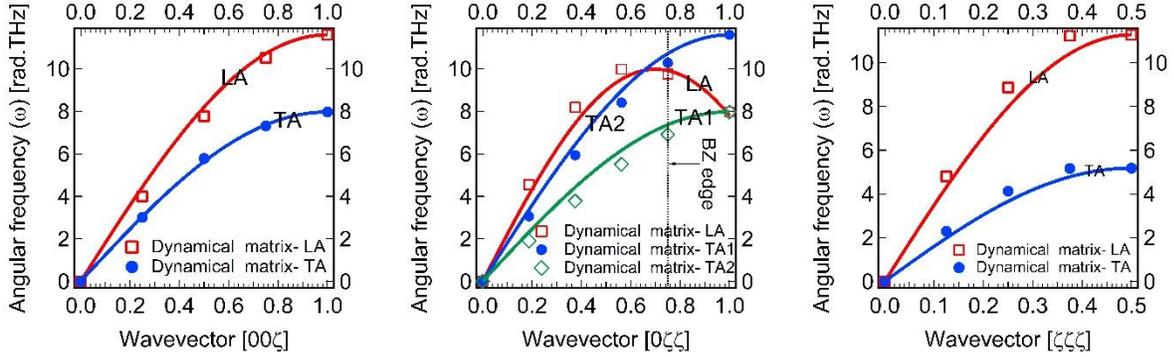

**FIG. 1. Phonon dispersion curves in the three high symmetry crystallographic directions [001], [110], and [111] (solid lines) for the three polarization branches: Longitudinal Acoustic (LA), high energy Transverse Acoustic (TA2), and low energy Transverse Acoustic (TA1), along with the eigenvalues for selected points calculated by solving the secular determinant of the dynamical matrix of argon. $\zeta$ is the Cartesian component of wave vector in reciprocal lattice unit of $2\pi/a$, where $a$ is the lattice constant. The two transverse branches are degenerate in the [001] and [111] directions.**

By revisiting Eq. (1) we realize that it is a system of coupled linear equations. One of the traditional ways to overcome this difficulty is the relaxation time approximation, under which Eq. (1) becomes

$$\frac{\hbar \omega_{qs}}{k_B T^2} v_g^{qs} \cdot \nabla T \, \bar{n}_{qs}(\bar{n}_{qs}+1) = -\frac{n_{qs} - \bar{n}_{qs}}{\tau_{qs}}. \tag{3}$$

By employing the picture of self-scattering, the relaxation time $\tau_{qs}$ is used as a time-scale for non-equilibrium excitation of a phonon state to decay. Phenomenological models, e.g., Holand's

Page | 8

model, for relaxation time of mode $qs$ at temperature $T$ were developed using power functions in terms of frequency and temperature; for example, $\tau_{qs} \sim \omega_{qs}^n T^m$, with $n$ and $m$ being adjustable constants. Comparing the scattering terms in Eq. (3) and (1), an expression for the relaxation time can be obtained from perturbation theory,

$$\tau_{qs}^{-1} = \sum_{q's',q''s''} [\frac{\bar{P}_{qs,q's'}^{q''s''}}{\bar{n}_{qs}(\bar{n}_{qs}+1)} \frac{(\psi_{qs}+\psi_{q's'}-\psi_{q''s''})}{\psi_{qs}} + \frac{1}{2}\frac{\bar{P}_{qs}^{q's',q''s''}}{\bar{n}_{qs}(\bar{n}_{qs}+1)} \frac{(\psi_{qs}-\psi_{q's'}-\psi_{q''s''})}{\psi_{qs}}]. \tag{4}$$

To evaluate the relaxation time in Eq. (4) the deviational terms ($\psi$'s) should be found. In this regard, different models are available with different levels of accuracy for modeling the correlation between the relaxations of different modes (the collective relaxation). The simplest is the Single Mode Relaxation Time (SMRT) approximation, which ignores the correlation completely and retains only the diagonal elements of the phonon collision operator. Under SMRT approximation, $\psi_{q's'} = \psi_{q''s''} = 0$ and Eq. (4) reduces to

$$\left(\tau_{qs}^{SMRT}\right)^{-1} = \sum_{q's',q''s''} [\frac{\bar{P}_{qs,q's'}^{q''s''}}{\bar{n}_{qs}(\bar{n}_{qs}+1)} + \frac{1}{2}\frac{\bar{P}_{qs}^{q's',q''s''}}{\bar{n}_{qs}(\bar{n}_{qs}+1)}]. \tag{5}$$

Other techniques include variational methods and the iterative method.[7,36] By using Eq. (3) and the linearized form of the phonon occupation number, $\psi$'s can be defined. Plugging this into Eq. (4) yields

$$\tau_{qs} = \tau_{qs}^{SMRT}(1+\Delta_{qs}), \tag{6}$$

where

$$\Delta_{qs} = \sum_{q's',q''s''} \left[ \frac{\bar{P}_{qs,q's'}^{q''s''}}{\bar{n}_{qs}(\bar{n}_{qs}+1)} \frac{(\tau_{q''s''} \upsilon_{g\|\nabla T}^{q''s''} \omega_{q''s''} - \tau_{q's'} \upsilon_{g\|\nabla T}^{q's'} \omega_{q's'})}{\upsilon_{g\|\nabla T}^{qs} \omega_{qs}} \right.$$
$$\left. + \frac{1}{2} \frac{\bar{P}_{qs}^{q's',q''s''}}{\bar{n}_{qs}(\bar{n}_{qs}+1)} \frac{(\tau_{q''s''} \upsilon_{g\|\nabla T}^{q''s''} \omega_{q''s''} + \tau_{q's'} \upsilon_{g\|\nabla T}^{q's'} \omega_{q's'})}{\upsilon_{g\|\nabla T}^{qs} \omega_{qs}} \right]. \quad (6a)$$

In this expression, $\upsilon_{g\|\nabla T}^{qs}$ is the component of the group velocity in the direction of the applied temperature gradient. It should be borne in mind that the magnitude of the temperature gradient appears in the numerator and the denominator, so they cancel out. However, unlike SMRT approximation, the value of the relaxation time will depend on the direction of the applied temperature gradient, due to the anisotropy in group velocities and phonon focusing.[7,38] This effect can cause anisotropy in thermal conductivity even for materials with cubic crystal system. This anisotropy should be understood as a consequence of the dependence of the steady state phonon occupation number for the same normal mode on the direction of the applied temperature gradient within the irreducible Brillouin zone in a nonlinear fashion, due to the coupling with the displacement of phonon occupation number of other phonon normal modes from equilibrium distribution, and not because of the ballistic mode of transport. The thermal conductivity is defined in this case as the linear proportionality constant between the applied temperature gradient and the conjugate heat flux, and second-rank tensor will not be appropriate to represent the conductivity macroscopically. By solving Eq. (6) iteratively, relaxation times are calculated. In this regard, we used a fixed point iteration scheme. Finally, from phonon kinetic theory, based on treating phonon system as gas of bosons occupying the crystal lattice, an expression for lattice thermal conductivity, $k$, of spatially homogenous system can be derived by substituting directly in the linearized BTE under relaxation time approximation. For cubical systems, this is given by

$$k = \frac{k_B}{3N_o\Omega} \sum_{qs} \left(\frac{\hbar\omega_{qs}}{k_B T}\right)^2 \bar{n}_{qs}(\bar{n}_{qs}+1)\tau_{qs}\left(\upsilon_g^{qs}\right)^2. \tag{7}$$

In our computation of Eqs. (5), (6), and (7) for argon, we treated the actual shape of the Face Centered Cubic (FCC) Brillouin zone (truncated octahedron) and considered all different branches of phonon polarization (phonon bands). Point group symmetry properties of FCC is exploited to reduce the computational domain to only 1/48 of Brillouin zone volume, which is called the irreducible Brillouin zone. A simple cubic mesh is used to tessellate the Brillouin zone and special q-points were taken as sample points. Given that phonon normal modes are uniformly distributed over the Brillouin zone, under cyclic condition, mode density weighted average (which is proportional to the fractional volume of the mesh elements) is used for all Brillouin zone sums. Accordingly, the complete Brillouin zone sum was replaced by partial zone sum using appropriate weighting factor. This discretization and summation procedure provides an efficient and fast convergent scheme, cf. Ref. 10. The values for lattice constant at 0 K (a = 0.53 nm) and the macroscopic Grüneisen parameter ($\gamma$ = 2.5) were taken from Ref. 28. For FCC, the primitive cell volume $\Omega$ is equal to one fourth of the conventional unit cell volume ($\Omega = a^3/4$). Moreover, the theoretical density is used in all simulations, for the sake of consistency.

To represent the cubic anisotropy of dispersion curves (without the need to solve the dynamical matrix for points in general directions), a direct linear interpolation scheme, over the irreducible wedge of the Brillouin zone, between dispersion data in the three high symmetry directions was implemented. For the sake of brevity, we will denote this dispersion model as FANISO model (short for Fully ANISOtropic). Group velocities are calculated using the first derivative of the analytical functions used to generate the dispersion curves. In addition, group velocities are postulated to be in the same direction of the wavevectors, which sounds appropriate for cubic crystal under quasi-continuum approximation, however a small angle between the two vectors may arise in a more rigorous treatment. More details on the numerical scheme will be addressed elsewhere[39].

Particular attention should be paid in handling Dirac delta function in discrete summation in Eqs. (2a) and (2b). In general, two main techniques have been sought in the literature to handle this, either to regularize the delta function using an approximate closed form (which integrates to unity), or to transform the discrete summation to a continuous integration. Although the second method handles the delta function in an exact way in terms of group velocity, a difficulty arises due to the need to determine the surfaces of constant energy ($S_\perp$), which usually introduces another source of approximation. This takes place through the transformation of the BZ sum/integration from the reciprocal space to the energy domain via: $d^3q = d|q| \cdot d^2q_\perp \simeq d\Delta|q| \cdot dS_\perp \simeq \frac{d\Delta|q|}{d\Delta\omega} d\Delta\omega \cdot dS_\perp = \upsilon_g d\Delta\omega \cdot dS_\perp$, so that performing the integration on Dirac delta function (defined in energy domain) is possible, or vice versa. When the surface of constant energy is hypothesized to be a plane normal to the group velocity (a common practice in most of the studies followed this approach),[26,36] the two expressions become equivalent.

For the regularization of the delta function method, depending on the range of the approximate function used, the conservation of energy can be treated implicitly or explicitly. For example, using a rectangular function, unit pulse, suitable under narrow resonance approximation, energy conservation rule is enforced explicitly by filtering out all phonon triplets that does not meet the specified criteria for energy conservation, while using the same weight for all triplets passing the energy filter. This is in contrast to other extended representations using continuous distribution functions (for example, Gaussian or Lorentz distribution functions) that treat energy conservation rules implicitly by accepting a weighted contribution from all phonon triplets. By close examination of the correspondence between the regularization approach and the integration approach, it can be observed that the width, and hence the height, of the delta function should not be left arbitrary. This is very critical, as it affects the effective number of phonon triplets that can pass the energy conservation test, which represents the available phase space for 3-phonon processes, see for example Ref. 40. So, the use of distribution function with a finite width should not be considered as a matter of approximation, it is substantiated on physical ground to allow the interaction between the otherwise non-interacting phonon normal modes. Unlike many other studies that used the width as an adjustable parameter, we here fix it on the basis of the aforementioned correspondence argument. The criterion is to fix the height such that the amplitude of the function at the root (i.e., $\omega_i = \omega_0$) is equivalent to the integral transform. To do so, the energy spacing of a mesh element $i$ ($\Delta_{\omega_i}$) was calculated by finding the energy difference between the farthest two surfaces of constant energy within the mesh element in the direction of maximum energy change. The connection between the delta function height and the mesh density is obvious through energy bin spacing dependence on the mesh density, which becomes mesh density independent when the Brillouin zone sum is carried out in energy domain. In our implementation, we tested Gaussian distribution function, Lorenz distribution, and the rectangular function. They take respectively the form:

$$\delta(\omega_i - \omega_0) = \frac{1}{\sqrt{2\pi}\sigma} e^{\frac{(\omega_i - \omega_0)^2}{2\sigma^2}}, \qquad \sigma = \frac{\Delta_{\omega_i}}{\sqrt{2\pi}}, \tag{8a}$$

$$\delta(\omega_i - \omega_0) = \frac{1}{\pi} \frac{\varepsilon}{(\omega_i - \omega_0)^2 + \varepsilon^2}, \qquad \varepsilon = \frac{\Delta_{\omega_i}}{\pi}, \tag{8b}$$

and

$$\delta(\omega_i - \omega_0) = \begin{cases} \dfrac{1}{\Delta_{\omega_i}}, & |\omega_i - \omega_0| \leq \dfrac{\Delta_{\omega_i}}{2} \\ 0, & \text{otherwise} \end{cases}. \tag{8c}$$

In Sec. III, we demonstrate the impact of the different representations of Dirac delta function presented here on phonon Density of States (DOS) and the thermal conductivity profile of solid argon over the entire temperature range (2 – 80 K). Furthermore, our calculated thermal conductivity using Lorenz distribution for two models of polarizations are compared with experimental data. To investigate the impact of the off-diagonal elements of phonon collision operator on thermal conductivity prediction, we compare our results using the iterative solver of BTE with SMRT approximation prediction for two different cases. The first case considers normal and umklapp processes as two resistive processes without making any distinction, while the second case ignores completely any role played by normal processes. In addition to the total lattice thermal conductivity, the individual contributions of individual phonon branches to thermal conductivity and the anisotropy of thermal conductivity (when temperature gradient is applied in different directions) are illustrated.

### III. RESULTS AND DISCUSSION

To quantitatively investigate the effect of considering energy conservation rule in statistical average sense on the intrinsic lattice thermal conductivity prediction due to 3-phonon processes, it is constructive to start by examining closely how the functional representation of energy broadening due to crystal anharmonicity affects phonon DOS. Fig. 2 shows phonon (DOS) using

Gaussian distribution, Lorentz distribution, and the rectangular approximation of the regularized Dirac delta. The discrepancy in DOS among these approximations occurs at the two extremes of the spectrum, where several orders of magnitude difference can be observed. This is attributed to the heavy-tailed nature of Lorentz distribution which enhances the DOS in the short and long wave limits appreciably. That is, the differences in the effective phonon DOS, within the employed energy continuum approximation, are mainly attributed to the sum of small contributions of energy-distant modes. As will be shown shortly, this helps in providing insight on the impact of the approximation that three-phonon processes are of elastic type, by assessing different levels of strictness in applying energy conservation rule.

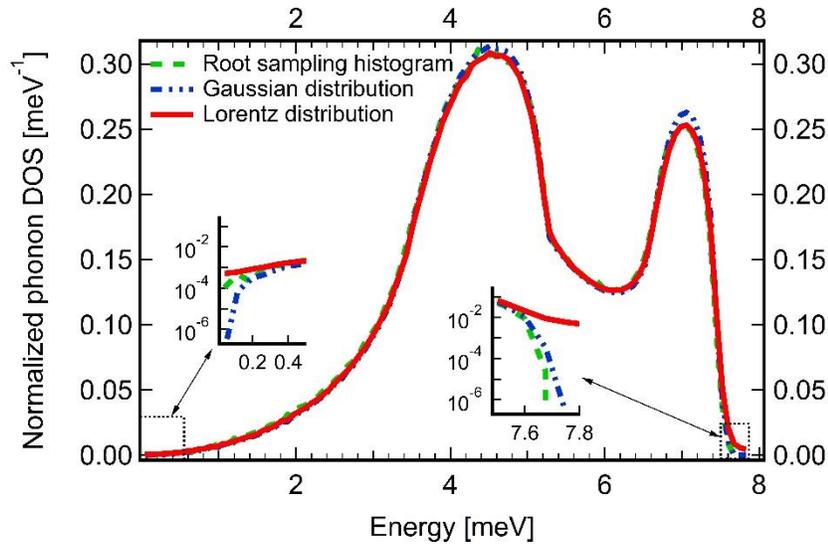

FIG. 2. Single mode phonon Density of States (DOS) using FANISO dispersion model and different representation for the Dirac delta function. The two insets magnify the behavior at the two extremes of the spectrum.

As we discussed in Sec. II, the width of regularized Dirac delta function controls the number of the available decay channels for a given phonon normal mode, and hence determines its scattering strength. By making a connection to the phonon lineshape to represent the statistical form of Dirac Delta function, this can be understood in some sense as a measure of the

Page | 15

probability density function for a given mode to have momentarily an energy (due to crystal anharmonicity) that can differ from the well-defined harmonic eigenenergy. Of course, in the absence of crystal anharmonicity this lineshape is represented by sharp peak with zero width representing a discrete eigenenergy. Another way to think about it is by recalling that the scattering in 3-phonon processes is inelastic type.[7] From both experiment and theory, we know that phonon normal mode self-energy exhibits the shape of a Lorentz distribution, and the first-order transition rates between pure harmonic phonon states can be directly extracted from phonon linewidth.[10,11] However, Dirac delta function has been mostly treated from a numerical rather than a physical perspective, not to mention the wide use of the integral transform method, which applies a strict energy conservation criterion on the available decay channels in three-phonon scattering phase space. This, in turn, excludes the contributions of the heavy tailed portion of Lorentz distribution to scattering strength, by limiting the uncertainty in the energy of a given phonon mode to a small value (i.e., assuming elastic scattering). Accordingly, treating 3-phonon interaction as an elastic scattering is a matter of approximation that needs to be assessed, and considering the actual shape of phonon energy broadening should be more reliable. So the next question should be about the impact of this particular shape on the thermal conductivity. In this regard, it is useful to mention that Turney et al.[11] used the same physical argument in their selection of the Lorentzian distribution to represent Dirac delta in the calculation of phonon relaxation times. The main difference in their case was the criterion they applied to determine the width of their resonance. They used individual phonon triplet linewidth, which is inversely proportional to the relaxation time itself. By doing so, Turney et al.[11] constructed what they called self-consistent loop to find the width of Dirac delta function in terms of the relaxation times iteratively. Since we are using energy continuum approximation, this discussion does not apply to individual normal modes. It applies to ensemble of phonon states having energies within increment *dE* about *E*, and the scattering, after the discretization of Brillouin zone, is ascribed to an effective resonance representative of the sum of the imaginary components of the self-energy of these individual phonon modes (individual resonances). Accordingly, in contrast to what Turney et al.[11] did, the correspondence argument, which is discussed in the previous section that defines the delta function as a function of energy mesh spacing, is employed here. It is worth to mention here that in our treatment, unlike the work of Turney et al., we use only harmonic properties to evaluate the width of the distribution, which depends on the group velocity and q-mesh cell spacing $\Delta_{q_i}$ through the relation: $\Delta_{\omega_i} \approx \upsilon_{g_i} \Delta_{q_i}$. The criterion we are using in the present study to assess the impact of different possible representation of Dirac delta function is based on investigating the predicted thermal conductivity profile as a function of temperature.

With their low or zero group velocity, the contribution of modes at Brillouin zone edge to thermal conductivity is negligible, and attention should be focused on the long-wave modes toward the center. Fig. 3 provides the answer. Strikingly, when Lorentz distribution is used a peak in the thermal conductivity is obtained, below which the thermal conductivity decreases toward zero at 0 K. This is in contrast to the exponential increase at low temperature when Gaussian distribution was used, and is independent of how the off-diagonal elements of the phonon collision kernel was treated. As it can be evidenced from the similar pattern obtained when both SMRT approximation and the iterative scheme were invoked with those two distribution functions. It should be emphasized here that both Gaussian and Lorentzian distributions yield the same value for 3-phonon scattering phase space and no excess scattering phase space is achieved in the case of the Lorentzian as compared to the Gaussian. The discrepancy in conductivity behavior at low temperature between these two distributions is related to the difference in the weight of the contribution of individual phonon triplets, with different energies and population, to the scattering strength. Furthermore, by truncating the range of the two distribution functions at the limit of their width (similar to Christen and Pollack treatment),[9] so that they have nonzero value only for energies that belong to the same energy bin where the root of the delta function is located, the characteristic behavior of the Lorentz distribution was lost. The results demonstrate that the conductivity peak arises mainly due to the tail portion of the distribution, which makes possible the interaction between the sparsely populated intermediate energy states at low temperature (with large wavevectors) and the densely populated long-wave modes, thus activating umklapp processes without the need to consider higher order phonon processes[35] or multi-step interactions[6] to rationalize the presence of the peak in the temperature dependence of conductivity.

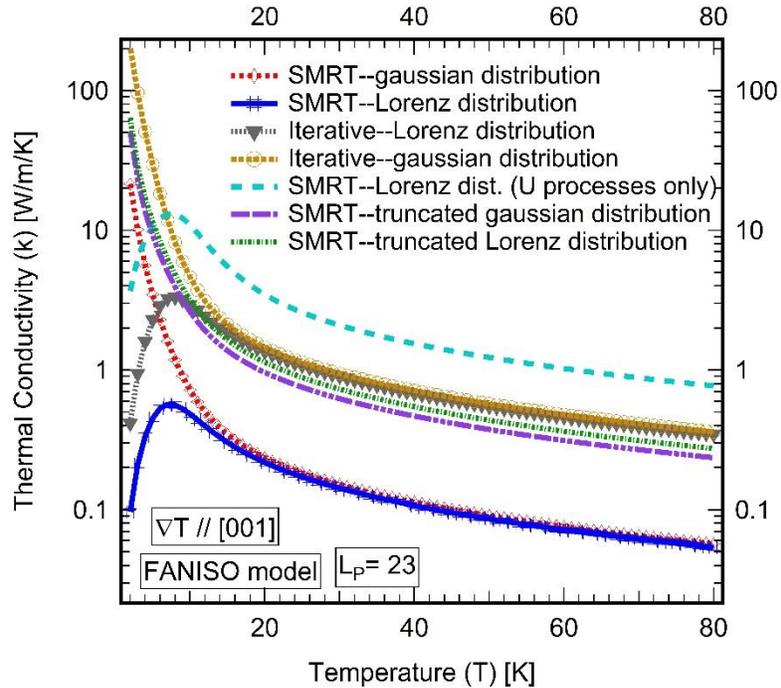

**FIG. 3.** Calculated thermal conductivity profile for different treatments of Dirac delta function, using FANISO dispersion model and when the temperature gradient is applied towards [001] direction ($L_P$ is the number of sample points along [001] crystallographic direction used in the simulations).

At high temperature this effect is unimportant, as can be seen from the coincidence of the two curves using Lorentz and Gaussian distributions. This is because long-wave modes are the dominant heat carriers at low temperature, but their relative contributions decrease remarkably at high temperature, as they become overpopulated. Another way to interpret it is that this coincidence at high temperature is as a direct implication of central limit theorem, as the number of scattering events in this limit is high. In agreement with the expectations from the heavy-tailed property of Lorentz distribution, the thermal conductivity values for truncated Lorentz distribution is higher than the truncated Gaussian distribution, because of the smaller value of the three-phonon scattering phase space in the first case.

For the assessment of the importance of the contribution of the off-diagonal elements of phonon collision operator (correlation effects) to thermal conductivity, results based on SMRT approximation (using both normal and umklapp processes) as well as umklapp processes alone are also demonstrated in Fig. 3, using Lorentz distribution for Dirac delta. It is obvious that SMRT approximation prediction of conductivity is lower than the iterative method prediction (and the experimental data, as will be shown later) by an order of magnitude. This is a direct manifestation that ignoring completely the non-resistive nature of normal processes is not adequate. On the other hand, by considering umklapp processes alone, and disregarding the interplay between normal and umklapp processes in lattice thermal resistivity, we overestimate the thermal conductivity by an order of magnitude. This overestimation is even true in the high temperature regime, where it is well-known that phonon-phonon interactions are the dominant scattering mechanism. It is not unusual in computational studies for lattice thermal conductivity prediction to apply inappropriately truncated distributions, treat the width of delta function as an adjustable parameter, and/or considering only umklapp processes to get values close to experiment by underestimating the scattering strength. All of these practices are unreliable and work more or less in an uncontrolled manner. Consequently, the determination of the coupling terms that appears in the collision kernel of the linearized form of BTE and the Lorentz distribution should be always sought for a more reliable thermal conductivity prediction.

It is worth mentioning that the impact of the mesh density and the number of iterations on the simulation results was investigated and temperature dependence was observed. Although the converged results are reported here, the conducted convergence study itself will be presented elsewhere[39]. In all the results presented here, a total of 1299 sample points, distributed uniformly over the irreducible Brillouin zone, were used. This corresponds to having 23 sample points along [001] crystallographic direction. In addition, the temperature gradient was taken to be parallel to [001] crystallographic direction.

As highlighted earlier, our results recover the right temperature dependence in the low ($T^2$) and high ($T^{-1}$) temperature limits, the peak about 8 K, and compares well with experiment, which is evidenced in Fig. 4. In this figure, the calculated thermal conductivity using our computational model (which employs Lorentz distribution, the iterative scheme for solving BTE, and our FANISO model for dispersion) is benchmarked against available experimental data for argon from several references. The results suggest that 3-phonon processes are the dominant scattering mechanism over the whole temperature range in argon. To investigate the effect of polarization type, which is commonly ignored under continuum approximation and replaced by phonon bands, we compare the standard model in the literature for the continuum representation of crystal anharmonicity, which considers randomly oriented eigenvectors (thus no distinction is made between longitudinal modes and transverse modes, based on the relative direction between the wavevector and polarization eigenvector), with what designated as "noTTT" model (short for no Transverse Transvers Transverse phonon triplet interactions). In the second model we exclude the possibility of interaction, if all of the three normal modes of a given phonon triplet belong to transverse branches (in similarity with pure transverse modes picture, where the polarization eigenvector is perpendicular to the wavevector). This produces values of thermal conductivity that are higher than our standard model, but still fall within the experimental data.

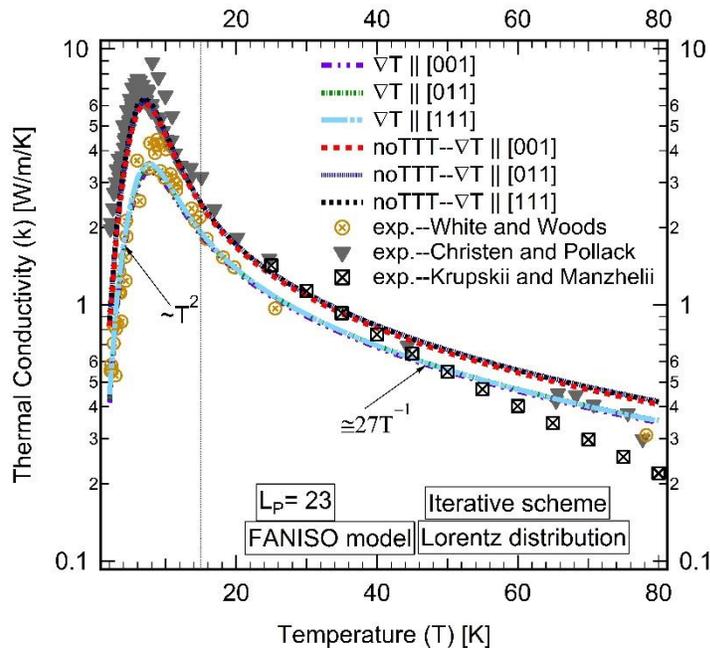

**FIG. 4. Simulated thermal conductivity, when the temperature gradient is applied towards the three high symmetry directions [001], [110], and [111] respectively, for the standard phonon band model (polarization-type independent), as well as "noTTT" model that prohibits interaction between phonon triplets if all of them have transverse polarization type, alongside with experimentally measured values for solid argon from several references.**

Several improvements could be supplemented to the standard model used here in this study including: a) accounting for the temperature dependence of the dispersion curves (using for example Quasi-harmonic approximation, or phonon self-consistent method), b) using more

Page | 21

accurate interatomic potential, c) seeking first-principle methods for harmonic and anharmonic interatomic force constants determination, or d) considering the mode-dependent nature of crystal anharmonicity (using, for example, mode-specific Grüneisen parameter). These improvements, however, will not change our conclusion regarding the impact of the Lorentz distribution on prediction of low temperature conductivity; the impact will only be a marginal refinement of results.

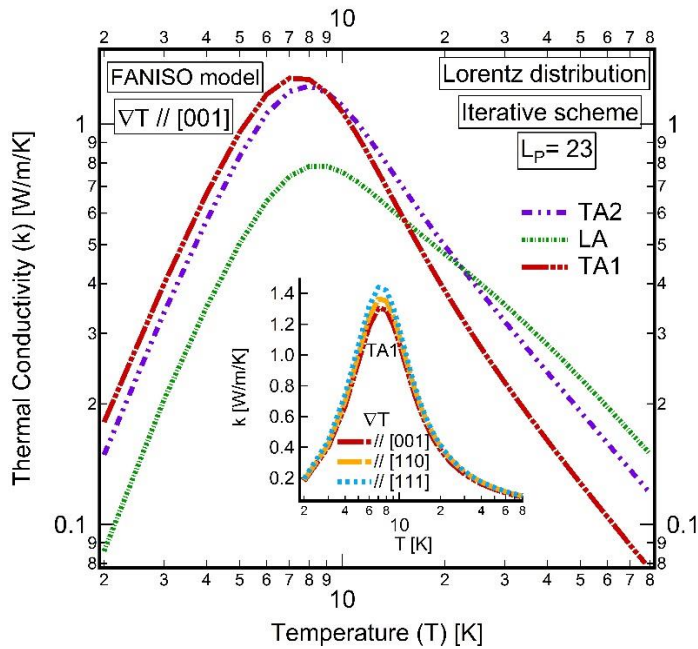

**FIG. 5. Temperature behavior of individual contributions of Longitudinal and Transverse Acoustic branches (LA, TA1, and TA2) to the lattice thermal conductivity for the standard model (when the temperature gradient is applied towards [001] direction), the inset shows anisotropy in TA1 branch, by changing the direction of the applied temperature gradient**.

Fig. 5 demonstrates the dominant role played by the low energy Transverse Acoustic (TA1) branch in thermal conductivity at low temperature. At the other extreme, the Longitudinal Acoustic (LA) branch contribution surpasses the separate contributions of the high energy Transverse Acoustic (TA2) and TA1 branches. In addition, three peaks at different temperatures are observed. Roughly speaking, the higher the energy of Brillouin zone edge of a given branch, the higher the temperature at which the peak contribution to thermal conductivity is located, and the peak tends to have lower value. By changing the direction of the applied temperature gradient between the three high symmetry crystallographic directions, anisotropy in thermal conductivity is predicted from our iterative scheme, as aforementioned, with maximum difference of 8% at the peak. Although this effect is persistent, it decays to very small differences above 10 K. As would be expected from the phonon focusing effect, the highest value for thermal conductivity was achieved when the temperature gradient was taken to be in the [111] crystallographic direction, since the edge energy of TA1 branch is the lowest in this direction (see Fig. 1). The inset of Fig. 5 clarifies this observation.

## IV. CONCLUDING REMARKS

To sum up, the critical importance of adopting Lorentz distribution to represent Dirac delta function in the calculation of the intrinsic lattice thermal conductivity at low temperature using Fermi golden rule was demonstrated. This helped us to get finite values for thermal conductivity at low temperature with a conductivity peak at the appropriate place. By utilizing macroscopic thermodynamic Grüneisen parameter, FANISO dispersion model for cubic anisotropy, and the iterative scheme to solve the linearized BTE, experimental thermal conductivity of FCC argon was fairly reproduced over the whole temperature range (2 – 80 K) by the sole use of 3-phonon processes. Remarkably, this evidences that phonon-phonon interaction mechanisms are effective over the entire temperature range including low temperature, which is contrary to the common consensus in the literature. In addition, anisotropy of thermal conductivity was captured. The current results indicate that the collective nature of phonon modes relaxation is critical for the right determination of the order of magnitude of thermal conductivity. Accordingly, the widely used SMRT approximation is not suitable for the case of solid argon.

ACKNOWLEDGEMENTS

This material is partly based upon work supported as part of the Center for Materials Science of Nuclear Fuel, an Energy Frontier Research Center funded by the U.S. Department of Energy, Office of Sciences, Office of Basic Energy Sciences under award number FWP 1356, through subcontract number 00122223 at Purdue University, and partly by Idaho National Laboratory through a subcontract titled 'Microstructure Evolution in $UO_2$' at Purdue University.

[1] G. P. Srivastava, Reports Prog. Phys. **78**, 026501 (2015).

[2] D. G. Cahill, P. V. Braun, G. Chen, D. R. Clarke, S. Fan, K. E. Goodson, P. Keblinski, W. P. King, G. D. Mahan, A. Majumdar, H. J. Maris, S. R. Phillpot, E. Pop, and L. Shi, Appl. Phys. Rev. **1**, 011305 (2014).

[3] P. G. Klemens, Solid State Phys. - Adv. Res. Appl. **7**, 1 (1958).

[4] R. Berman, Cryogenics (Guildf). **5**, 297 (1965).

[5] J. M. Ziman, *Electrons and Phonons* (Oxford University Press, London, 1960).

[6] Y.-J. Han and P. G. Klemens, Phys. Rev. B **48**, 6033 (1993).

[7] G. P. Srivastava, *The Physics of Phonons* (Taylor and Francis, New York, 1990).

[8] A. A. Maznev and O. B. Wright, Am. J. Phys. **82**, 1062 (2014).

[9] D. K. Christen and G. L. Pollack, Phys. Rev. B **12**, 3380 (1975).

[10] D. C. Wallace, *Thermodynamics of Crystals* (Dover Publications, INC, New York, 1998).

[11] J. E. Turney, E. S. Landry, A. J. H. McGaughey, and C. H. Amon, Phys. Rev. B - Condens. Matter Mater. Phys. **79**, 1 (2009).

[12] A. Togo, L. Chaput, and I. Tanaka, Phys. Rev. B - Condens. Matter Mater. Phys. **91**, 94306 (2015).

[13] A. Berne, G. Boato, and M. De Paz, Nuovo Cim. B **46**, 182 (1966).

[14] H. Egger, M. Gsänger, E. Lüscher, and B. Dorner, Phys. Lett. A **28**, 433 (1968).

[15] I. N. Krupskii and V. G. Manzhelii, Sov. Phys. JETP - USSR **28**, 1097 (1969).

[16] I. N. Krupskii and V. G. Manzhely, Phys. Stat. Sol. **24**, K53 (1967).

[17] G. K. White and S. B. Woods, Philos. Mag. **3**, 785 (1958).

[18] G. K. White and S. B. Woods, Nature **177**, 851 (1956).

[19] F. Clayton and D. N. Batchelder, J. Phys. C, Solid State Phys. **6**, 1213 (1973).

[20] D. E. Daney, Cryogenics (Guildf). **11**, 290 (1971).

[21] H. R. Glyde and M. G. Smoes, Phys. Rev. B **22**, 6391 (1980).

[22] I. J. Gupta and S. K. Trikha, Phys. Stat. Sol. **80**, 353 (1977).

[23] I. J. Gupta and S. K. Trikha, Phys. Stat. Sol. **84**, K95 (1977).

[24] N. P. Gupta, J. Solid State Chem. **5**, 477 (1972).

[25] N. P. Gupta and P. K. Garg, Ann. Phys. (N. Y). **95**, 281 (1975).

[26] C. Julian, Phys. Rev. **137**, A128 (1965).

[27] O. N. Bedoya-Martínez, J.-L. Barrat, and D. Rodney, Phys. Rev. B **89**, 014303 (2014).

[28]G. K. Horton and J. W. Leech, Proc. Phys. Soc. **82**, 816 (1963).

[29]G. Niklasson, Phys. Kondens. Mater. **14**, 138 (1972).

[30]H. Kaburaki, J. Li, S. Yip, and H. Kimizuka, J. Appl. Phys. **102**, 043514 (2007).

[31]M. Omini and A. Sparavigna, Philos. Mag. Part B **68**, 767 (1993).

[32]M. Omini and A. Sparavigna, Phys. Rev. B **53**, 9064 (1996).

[33]A. J. H. McGaughey and M. Kaviany, Phys. Rev. B **69**, 1 (2004).

[34]A. J. H. McGaughey and M. Kaviany, Int. J. Heat Mass Transf. **47**, 1783 (2004).

[35]T. Feng and X. Ruan, Phys. Rev. B **93**, 045202 (2016).

[36]A. Chernatynskiy and S. R. Phillpot, Phys. Rev. B - Condens. Matter Mater. Phys. **82**, 1 (2010).

[37]D. A. Broido, M. Malorny, G. Birner, N. Mingo, and D. A. Stewart, Appl. Phys. Lett. **91**, 23 (2007).

[38]A. G. Every, W. Sachse, K. Y. Kim, and M. O. Thompson, Phys. Rev. Lett. **65**, 1446 (1990).

[39]A. Hamed, and A. El-Azab, "Quantitative assessment of perturbation theory-based lattice thermal conductivity models using quasi-continuum approximation", Manuscript Submitted for Publication (2017).

[40]W. Li, J. Carrete, N. a. Katcho, and N. Mingo, Comput. Phys. Commun. **185**, 1747 (2014).